\documentclass[smallabstract,smallcaptions]{dccpaper}

\usepackage{epsfig}
\usepackage{citesort}
\usepackage{amsmath}
\usepackage{amssymb}
\usepackage{color}
\usepackage{url}

\usepackage{graphicx}

\newlength{\figurewidth}
\newlength{\smallfigurewidth}

\begin{document}

\title{\large
\textbf{A conditional compression distance\\
that unveils insights of the genomic evolution}}

\author{%
Diogo Pratas and Armando J. Pinho\\[0.5em]
{\small\begin{minipage}{\linewidth}\begin{center}
\begin{tabular}{ccc}
\multicolumn{3}{c}{IEETA / Dept of Electronics, Telecommunications
and Informatics}\\
\multicolumn{3}{c}{University of Aveiro, 3810--193 Aveiro, Portugal}\\
\multicolumn{3}{c}{\url{pratas@ua.pt} --- \url{ap@ua.pt}}
\end{tabular}
\end{center}\end{minipage}}
}

\maketitle
\thispagestyle{empty}

\begin{abstract}
We describe a compression-based distance for genomic sequences.
Instead of using the usual conjoint information content, as in the
classical Normalized Compression Distance (NCD), it uses the conditional
information content. To compute this Normalized Conditional Compression
Distance (NCCD), we need a normal conditional compressor, that we
built using a mixture of static and dynamic finite-context models.
Using this approach, we measured chromosomal distances between
\textit{Hominidae} primates and also between \textit{Muroidea}
(rat and mouse), observing several insights of evolution that so
far have not been reported in the literature.
\end{abstract}

%
\Section{Introduction}
\label{sec:intro}
The high-throughput sequencing technologies are creating an avalanche of
genomic and metagenomic sequences, nonexistent a few years ago. We are
now able to computationally evaluate similarities, or their absence,
among species and across different regions of the same species, using
whole genomes.

Common biological approaches for determining distances, usually using
FISH techniques, are very expensive and time-consuming. Computational
approaches have emerged as an affordable, fast and automated process to
deal with this problem. Several computational distance metrics have been
proposed, where some of the most popular are the Hamming \cite{Hamming-1950a}
and Levenshtein \cite{Levenshtein-1966a} distances. The Hamming distance
can only be applied when the sequences are aligned with precision and
have the same size, requirements hardly found in large genomic sequences.
The Levenshtein distance explores transformations between the sequences,
namely insertions, deletions and substitutions. Although quite successful,
its computational time is prohibitive for large sequences (the fastest
known implementation runs with time complexity $O(n^2/\log n)$).

Compression-based approaches emerged as a natural way for measuring
distances, because, together with the appropriate decoder, the bitstream
produced by a lossless compression algorithm allows the reconstruction
of the original data and, therefore, can be seen as an upper bound of
the algorithmic entropy of the sequence. A compression-based distance
computes the distance between two objects using the number of bits
needed to describe one of them when a description of the other is
available, as well as the number of bits required to describe each of them.

Compression-based distances are founded on the Kolmogorov notion
of complexity, also known as algorithmic entropy, where $K(x)$
of a string $x$ is the length
of the shortest binary program $x^*$ that computes $x$ in an appropriate
universal Turing machine and halts \cite{Turing-1936a}. As such,
$K(x) = |x^*|$, the length of $x^*$, denotes the number of bits of
information from which $x$ can be computationally retrieved \cite{Li-2008b}.
The conditional Kolmogorov complexity, $K(x|y)$, denotes the 
length of the shortest binary program, in the universal prefix Turing
machine, that on input $y$ outputs $x$. A special case occurs
when $y$ is an empty string, $y = \lambda$, and hence $K(x|\lambda) = K(x)$.

Bennett introduced the information distance \cite{Bennett-1998a},
$E(x,y) = \max\{K(x|y),K(y|x)\}$, defined as the length of the shortest
binary program for the reference universal prefix Turing machine that with
input $x$ computes $y$, as well as with $y$ computes $x$. The normalized
version (NID \cite{Li-2004a}) of $E(x,y)$  is defined as
\begin{equation}\label{nid}
\mathrm{NID}(x,y) = \frac{\max\{K(x|y), K(y|x)\}}{\max\{K(x),K(y)\}},
\end{equation}
up to an additive logarithmic term. The normalized compression distance
(NCD) \cite{Cilibrasi-2005a} emerged to efficiently compute the NID, due
to the non-computability of $K$,
\begin{equation}\label{ncd}
\mathrm{NCD}(x,y) = \frac{C(x,y) - \min\{C(x),C(y)\}}{\max\{C(x),C(y)\}},
\end{equation}
up to an additive logarithmic term, where $C(x)$ and $C(y)$ represent,
respectively, the number of bits of the compressed version of $x$ and $y$,
and $C(x,y)$ the number of bits of the conjoint compression of $x$
and $y$ (usually, $x$ and $y$ are concatenated). Distances near one
indicate dissimilarity, while distances near zero indicate similarity.
It can be seen that for $\mathrm{NCD}(x,x) = 0$ to hold, then the
compressor needs to verify $C(x,x) \approx C(x)$, one of the most
important properties of a \textit{normal} compressor \cite{Cilibrasi-2005a}.

In this paper, we describe an admissible normalized compression
distance, relying on a conditional compressor, that builds an
internal model of the data using a mixture of static and dynamic
finite-context models (FCMs). We assess the metric and its inherent
parameterized compressor, and we present some results of chromosomal
distances between several large eukaryotic chromosomes, namely
\textit{Hominidae} primates and \textit{Muroidea}, confirming
several documented results and pointing out some undocumented
observations.

%
\Section{Proposed Approach}
\label{method}
A direct substitution of $K$ by $C$ in~(\ref{nid}) would require the
availability of compressors that are able to produce conditional
compression, i.e., $C(x|y)$ and $C(y|x)$. Most compressors do not have
this functionality and, therefore, the NCD avoids it by using suitable
manipulations of~(\ref{nid}) \cite{Cilibrasi-2005a}. Instead of
$C(x|y)$ and $C(y|x)$, a term corresponding to the conjoint compression
of $x$ and $y$, $C(x,y)$, was preferred.
Usually, this $C(x,y)$ term is interpreted as the compression of the
concatenation of $x$ and $y$, but, in fact, it could be any other form
of combination between $x$ and $y$. Concatenation is often used
because it is easy to obtain, but in fact its use may hamper the
efficiency of the measure \cite{Cebrian-2005a}.

To overcome this limitation, we propose use the direct form, to which
we call the Normalized Conditional Compression Distance (NCCD),
\begin{equation}\label{nccd}
\mathrm{NCCD}(x,y) = \frac{\max\{C(x|y), C(y|x)\}}{\max\{C(x),C(y)\}},
\end{equation}
where ``Conditional'' means that the compressor $C$ needs to be
able to perform conditional compression.

\SubSection{The conditional compressor}
We have built a NCCD compressor based on two model classes (we call
them ``static'' and ``dynamic''), each one composed of mixtures
of finite-context models (FCMs)
of several orders \cite{Bell-1990a,Salomon-2007a,Sayood-2012a}. To
compute $C(x|y)$, the compression is performed in two phases. In the
first phase, the static class of FCMs accumulates the counts regarding
the $y$ object. After the entire $y$ object was processed, the models
are kept frozen and, hence, the second phase starts. At this point, the $x$
object starts to be compressed using the static models computed during
the first phase, in cooperation with the set of FCMs of the dynamic
class, that dynamically accumulate the counts only from $x$.


The probability of each symbol is obtained by mixing the
probabilities provided by each FCM of the static and dynamic models,
using a weighted average, according to
\begin{equation}
P(x_{n+1}) = \sum_k P(x_{n+1} | x_{n-k+1..n})\;w_{k,n},
\end{equation}
where $w_{k,n}$ denotes the weight assigned to the finite-context model
$k$ and $\sum_k w_{k,n} = 1$. The conditional probabilities are given by
the estimator
\begin{equation} \label{eq:pe}
P(s | x_{n-k+1..n}) = \frac{C(s | x_{n-k+1..n}) + \alpha}
{C(x_{n-k+1..n}) + 4\alpha},
\end{equation}
where $C(s | x_{n-k+1..n})$ represents the number of times that, in the
past, symbol $s$ was found having $x_{n-k+1..n}$ as the conditioning
context and where $C(x_{n-k+1..n})$ is the total number of events that
has occurred so far in association with context $x_{n-k+1..n}$.

For stationary sources, we could compute weights such that
$ w_{k,n} = P(k | x_{1..n})$, i.e., according to the probability that 
model $k$ has generated the sequence until that point. In that case, 
we would get
\begin{equation}
w_{k,n} = P(k | x_{1..n}) \propto P(x_{1..n} | k) P(k),
\end{equation}
where $P(x_{1..n} | k)$ denotes the likelihood of sequence $x_{1..n}$
being generated by model $k$ and $P(k)$ denotes the prior probability
of model $k$. Assuming $P(k) = 1/K$, where $K$ denotes the total
number of FCMs, we obtain $w_{k,n} \propto P(x_{1..n} | k)$.
Calculating the logarithm we get
\begin{equation}
\log_2 P(x_{1..n} | k) = \log_2 \prod_{i=1}^n P(x_i | k, x_{1..i-1}) =
  \sum_{i=1}^n \log_2 P(x_i | k, x_{1..i-1}),
\end{equation}
which corresponds to the code length that would be required by model $k$
for representing the sequence $x_{1..n}$. It is, therefore, the accumulated
measure of the performance of model $k$ until instant $n$. However, since
the DNA sequences are not stationary, a good performance of a model in a
certain region of the sequence might not be attained in other regions.
Hence, the performance of the models have to be measured in the recent
past of the sequence, for example using a mechanism of progressive
forgetting of past measures. For that, we use the recursive relation
\begin{subequations}
\label{recursion}
\begin{gather}
\sum_{i=1}^n \log_2 P(x_i | k, x_{1..i-1}) = \\
  = \gamma\sum_{i=1}^{n-1} \log_2 P(x_i | k, x_{1..i-1}) +
  \log_2 P(x_n | k, x_{1..n-1}).
\end{gather}
\end{subequations}
This relation corresponds to a first-order recursive filter that, for
$\gamma \in [0,1)$, has a low-pass characteristic and an 
exponentially decaying impulse response. For additional information,
see, for example, \cite{Pratas-2011a,Pinho-2011a}.

\begin{figure}[ht]
\centerline{\includegraphics[width=10cm]{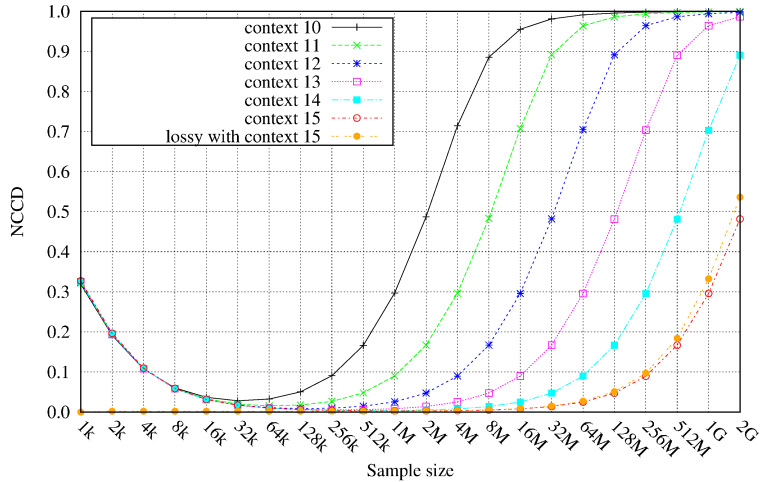}}
\caption{$\mathrm{NCCD}(x,x)$ value on uniformly distributed DNA
(synthetic) sequences with custom sizes, for several depths of the
highest order model. The ``\textit{lossy}'' curve shows the behavior
of NCCD when the best FCM is chosen for each base, corresponding
to a lower bound of the (non-reversible) compressor.}
\label{sizes}
\end{figure}

%
\Section{Parameterization and assessment}
\label{assess}
The parameters used in each compression measure must be kept constant, in
order to be used as a valid comparable metric between distances (otherwise
it will change the meaning of $C$). Accordingly, we have used a fixed setup
of five static FCMs and three dynamic FCMs, mixed using a set of weights
estimated with $\gamma = 0.9$. From our experience, we have verified
that $\gamma = 0.99$ maximizes the compression gain for bacterial genomes,
while for eukaryotic genomes $\gamma = 0.9$ seems to be the best choice.
The orders used for the static models were: 4, 6, 8, 10 and 15. For the
first four we used $\alpha = 1$ (Laplace estimator), whereas the one with
the highest order we used $\alpha = 0.001$. Usually, a small $\alpha$ is
important only for high orders (above ten). Moreover, the high order used
(15) ensures an admissible identity (i.e., $\mathrm{NCCD}(x,x) \approx 0$),
as Fig.~\ref{sizes} suggests.
The curve in Fig.~\ref{sizes} labeled ``\textit{lossy}'' corresponds
to using always the best FCM for each base and shows that the first
part of the curves is due to the adaptation of the method when not
enough data is present, suggesting that a very small sequence may
harm the identity property, also observed in very large sequences.
The latter drawback may be overcome using higher FCM orders, at the
cost of additional computational memory.

The three FCMs of the dynamic class have orders 4, 10 and 15, where the
first two rely on a Laplace probability estimator and the last one use
$\alpha = 0.05$. For the two deeper models, the inverted repeats are
also taken into account \cite{Pinho-2008a}. The maximum counters used
in each static model were, respectively, $2^9$, $2^{12}$, $2^{12}$.
This limitation acts also as a forgetting mechanism, because the counters
are divided by two when one of them reaches the maximum, decreasing the
importance of statistics collected in the far past. More information
regarding FCM parameterization can be obtained in
\cite{Pratas-2011a,Pinho-2011a,Pinho-2011e}.

\begin{figure}[ht]
\centerline{\includegraphics[width=10cm]{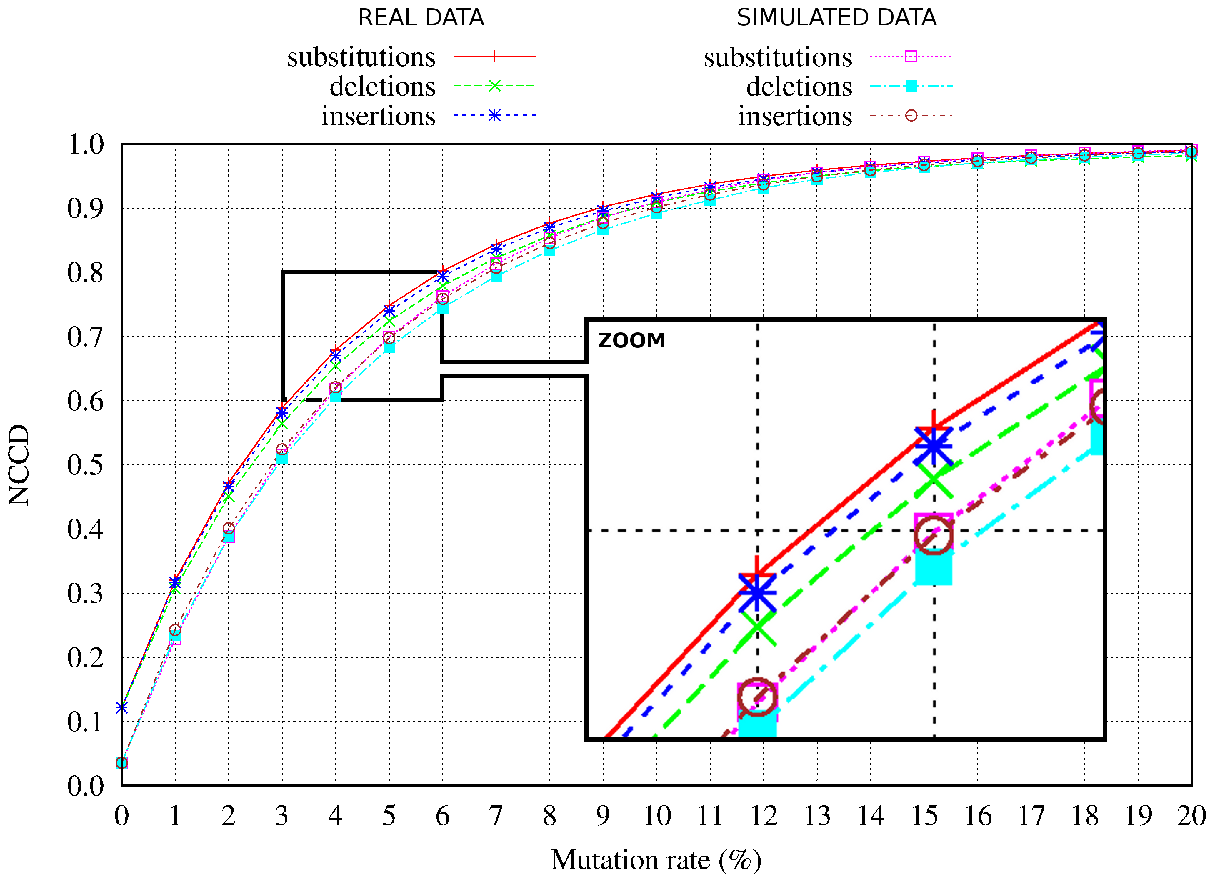}}
\centerline{\includegraphics[width=10cm]{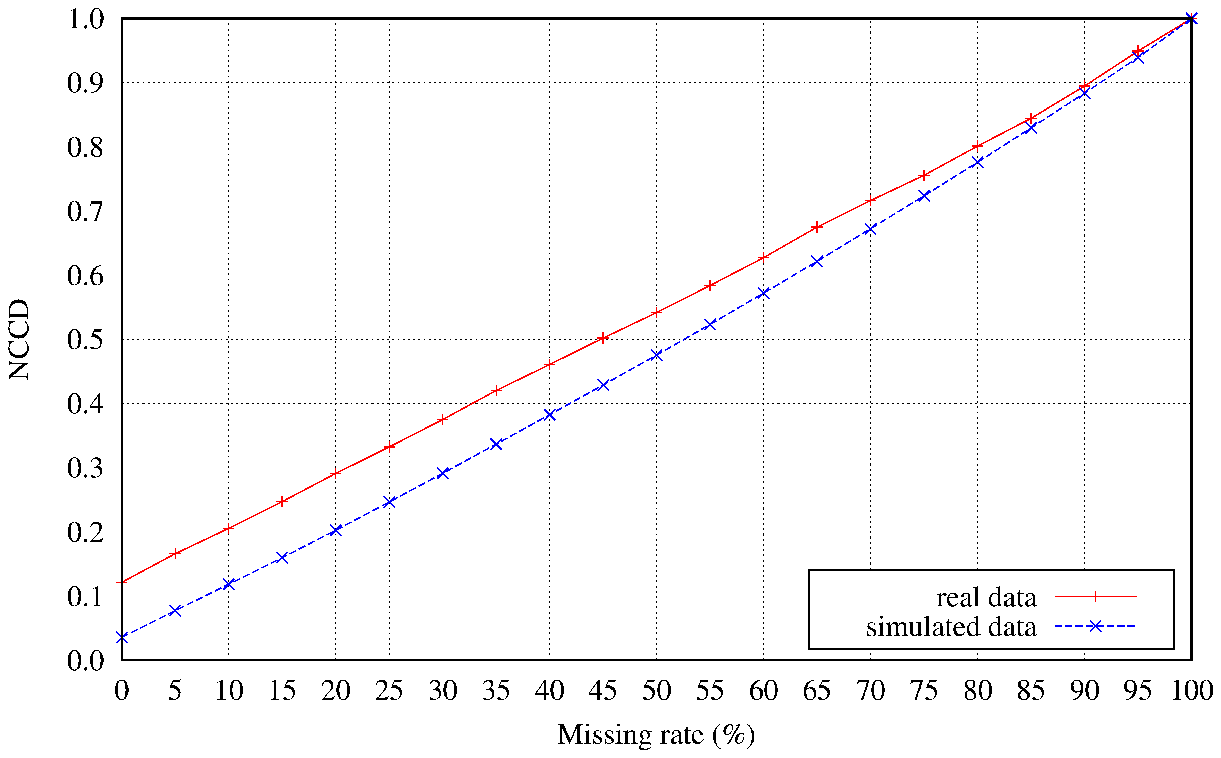}}
\caption{NCCD performance on synthetic and \textit{real} 50~MB of genomic
mutated data (top) and on progressive block missing data (bottom).}
\label{mutation}
\end{figure}

The DNA data sequences are products of sequencing techniques, which
have a sequencing quality, coverage and assembly technique associated 
\cite{Church-2011a}. Although these external factors may sometimes
constitute a problem, we believe that generally they are mitigated
by the compressor \cite{Cebrian-2007a}. Nevertheless, since we use a
metric based on conditionals targeting genomic sequences, we have
assessed the impact of uniformly distributed mutations, namely
substitutions, insertions and deletions, over 50~MB of \textit{real}
(first 50~MB of chromosome 1 from \textit{H. sapiens}) and synthetic
(simulated using XS from Exon \cite{Pratas-2012a}) genomic data, as
can be seen in the top graph of Fig.~\ref{mutation}. Substitutions
seem to be the most difficult mutation type to be handled by the
compressor, although only slightly, and, hence, by the NCCD. Although
it is clear that the method is still reporting reasonable distances
for sequences with 10\% of mutations, both for the \textit{real} and
synthetic sequences.

Finally, we have assessed the importance of sequence completeness using
progressive missing data, as the bottom graph of Fig.~\ref{mutation} depicts.
As expected, it is characterized by an approximately linear behavior.
However, there is a gap between the curves of the \textit{real} and
synthetic sequences, specially when there are lower missing rates.
This is due to the nature of the sequences, namely the self-similarity,
since the beginning of the \textit{real} sequence is composed by a
telomeric zone (highly-repetitive). On the other hand, the synthetic
sequence does not yield an exact zero of the NCCD when the missing rate
is zero, because it has been simulated with several approximately
repeating zones. This may be overcome with higher FCM orders, although
at the cost of more computer memory.

%
\Section{Results}
\label{results}
The data set is composed of six genomes (Table~\ref{dataset}),
downloaded from the NCBI website (\url{ftp://ftp.ncbi.nlm.nih.gov/genomes}).

\begin{table}\small
\caption{Data set used in the experiments. The number of expected
chromosome pairs for each species is represented by ``Exp'', while
``Missing'' is a nonexistence sequence and Mb represents the
approximated size in Mega bases.}
\begin{center}
\begin{tabular}{|r|r|r|r|r|}\hline
Organism & Build & Exp & Missing & Mb\\\hline\hline
\textit{Homo sapiens} & 37.p10 & 23 &- & 2,861 \\\hline
\textit{Pan troglodytes} & 2.1.4 & 24 & - & 2,756 \\\hline
\textit{Gorilla gorilla} & r100 & 24 & Y & 2,719 \\\hline
\textit{Pongo abelii} & 1.3 & 24 & Y & 3,028 \\\hline
\textit{Mus musculus} & 38.p1 & 20 & - & 2,716 \\\hline
\textit{Rattus norvegicus} & 5.1 & 21 & Y & 2,443 \\\hline
\end{tabular}
\end{center}
\label{dataset}
\end{table}

\begin{figure}[t]
\centerline{\includegraphics[width=15cm]{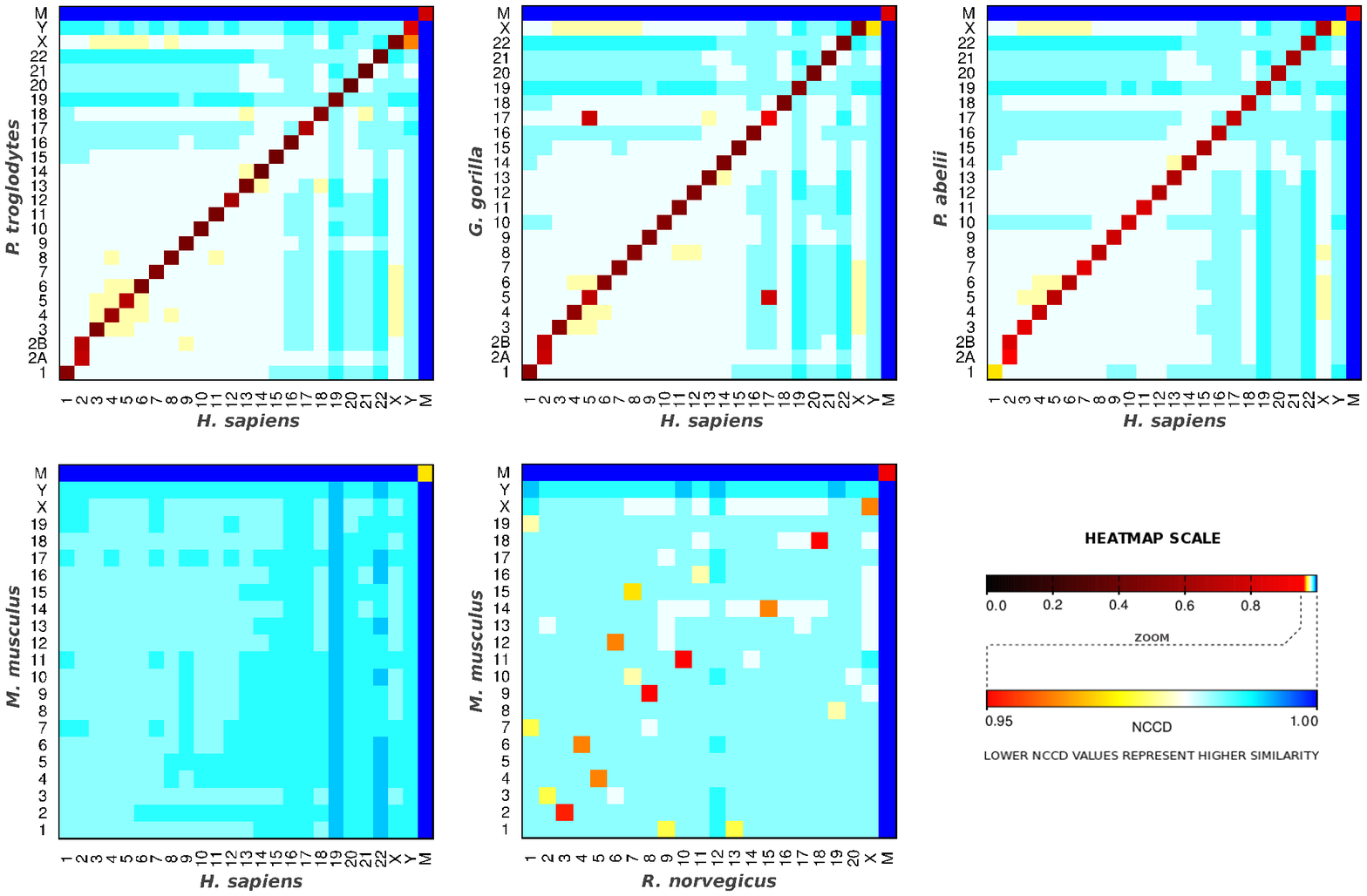}}
\vspace{0.5ex}
\caption{\textit{P. troglodytes}, \textit{G. gorilla}, \textit{P. abelii} and
\textit{M. musculus} inter-genomics chromosomal NCCD heatmaps, in relation to 
\textit{H. sapiens}, and \textit{M. musculus} in relation to
\textit{R. norvegicus}.}
\label{inter}
\end{figure}

Figure~\ref{inter} presents the inter-chromosomal NCCD distance heatmaps
relatively to \textit{H. sapiens} with the rest of the primates and
\textit{M. musculus}, and \textit{M. musculus} relatively to
\textit{R. norvegicus}, plotted in an \textit{all with all} scheme.
As can be seen, for all primate species there is a direct correlation
with the respective chromosomal number, with the exception
of chromosome~2 (related to 2A and 2B). This is justified by a presumed
chromosomal fusion in humans from previous ancestors \cite{Ijdo-1991a}.

Moreover, the human Y chromosome is highly related with the X chromosome
of other primate species, namely the \textit{P. troglodytes},
because the Y chromosome exchanged genetic information with X in the
recombination process \cite{Hughes-2010a}. Furthermore, there is a low
distance between chromosomes 5 and 17 of the \textit{G. gorilla} and
\textit{H. sapiens}, justified by a chromosomal translocation
\cite{Samonte-2002a}.

Relatively to \textit{M. musculus}, there is an obvious similarity with
\textit{R. norvegicus}, although smaller than in \textit{P. maniculatus} /
\textit{M. norvegicus} \cite{Ramsdell-2008a}. When compared with the
primates, no important similarities are found (at a genomic level), specially
in human chromosomes 19 and 22. Moreover, it seems that only the mithocondrial
sequences attain some level of similarity. Nevertheless, the
\textit{M. musculus} (MM) and \textit{R. norvegicus} (RN)
\textit{diagonal} is very dissipated for such a low distance depicted
in the mithocondrial sequence. In fact, only chromosomes (C) 18 and X
seem to be homologous (in the \textit{diagonal}). Subsequent analysis show 
strong similarity between MM C2 / RN C3, MM C9 / RN C8 and
MM C11 / RN C10, and considerable similarity between  MM C4 / RN C5,
MM C6 / RN C4, MM C12 / RN C6 and MM C14 / RN C15, without
detracting other important patterns.

\begin{figure}[t]
\centerline{\includegraphics[width=14cm]{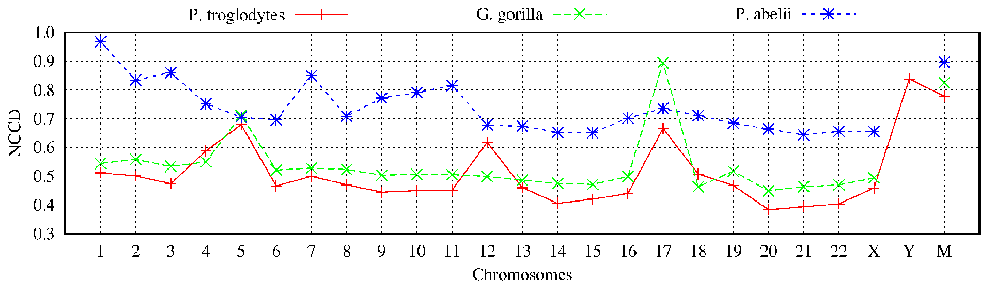}}
\vspace{0.5ex}
\caption{\textit{P. troglodytes}, \textit{G. gorilla} and 
\textit{P. abelii} related chromosomal NCCD values using 
\textit{H. sapiens} as reference.}
\label{diagonal}
\end{figure}

Figure~\ref{diagonal} presents the chromosomal distances of
\textit{P. troglodytes}, \textit{G. gorilla} and \textit{P. abelii} 
(chromosomes 2A and 2B have been concatenated) according to the
\textit{H. sapiens} chromosomes order. At glance, \textit{P. troglodytes}
has the lowest distance relatively to \textit{H. sapiens}, followed by
\textit{G. gorilla} and \textit{P. abelii}, respectively. Specifically,
the \textit{G. gorilla} chromosomes 5 and 17 have large distances
because of the previous mentioned translocation, while
\textit{P. abelii} seems to have a very different chromosome 1,
besides other relevant dissimilarities.

According to \cite{Mikkelsen-2005a}, besides the high divergence of
Y chromosome, there are several breakpoints in chromosomes 4, 5 and 12,
which were tested by fluorescence \textit{in situ} hybridization ({FISH})
in \textit{P. troglodytes}, using \textit{H. sapiens} as reference.
Figure~\ref{diagonal} reports the same dissimilarities, surprisingly
adding chromosome 17.

Finally, we have found that chromosomes 4, 12 and 18 of \textit{G. gorilla}
have lower distances to \textit{H. sapiens} than to the respective
\textit{P. troglodytes} chromosomes, while chromosomes 5 and 17 of
\textit{G. gorilla} have higher distances than those of \textit{P. abelii}.
Mitochondrial sequences, as expected, show that \textit{P. troglodytes}
is the nearest species to \textit{H. sapiens}, followed by the
\textit{G. gorilla} and \textit{P. abelii}.

%
\Section{Conclusion}
We have described a compressed-based metric for measuring distances between
genomic sequences, based on the conditional information content. This
approach requires a \textit{normal} conditional compressor, that we have
defined and assessed in this work. The compressor is constituted by a set
of multiple static and dynamic finite-context models, that cooperate under
a supervision mixture model. It is able to handle several types of
mutations, and hence rendering it a good candidate to study large
eukaryotic chromosomes. We have calculated chromosomal distances between
\textit{Hominidae} primates and also \textit{Muroidea} (rat and mouse)
rodents superfamily, attaining results that agree with several already
documented results, mainly using expensive and time-consuming FISH
approaches, but also unveiling undocumented ones.

\Section{Acknowledgements}
This work was supported in part by FEDER through the Operational Program 
Competitiveness Factors - COMPETE and by National Funds through FCT - 
Foundation for Science and Technology, in the context of the projects 
FCOMP-01-0124-FEDER-022682 (FCT reference PEst-C/EEI/UI0127/2011) and 
Incentivo/EEI/UI0127/2013.

\Section{References}
\bibliographystyle{IEEEtran}


\end{document}